\documentclass[11pt]{article}
\usepackage{hyperref}
\pdfoutput=1
\begin{document}
\title{Unsteady Flow and Control in Butterfly Take-off Flight of \\ Fluid Dynamics Videos}
\author{Haibo Dong, Chengyu Li, Zongxian Liang and Xiang Yun \\
\\\vspace{6pt} Department of Mechanical and Aerospace Engineering, \\ University of Virginia, Charlottesville, VA 22904, USA}
\maketitle

\begin{abstract}
In this work, high-resolution, high-speed videos of a Monarch butterfly (Danaus plexippus) in take-off flight were obtained using a photogrammetry system. Using a 3D subdivision surface reconstruction methodology, the butterfly's body/wing deformation and kinematics were modeled and reconstructed from those videos.  High fidelity simulations were then carried out in order to understand vortex formation in both near-field and far-field of butterfly wings and examine the associated aerodynamic performance. A Cartesian grid based sharp interface immersed boundary solver was used to handle such flows in all their complexity.
\end{abstract}

\end{document}